\documentclass[aps,prc,twocolumn,superscriptaddress,showpacs,showkeys,10pt,amsmath,amssymb]{revtex4-1}
\usepackage{graphicx}

\begin{document}

\title{Neutron pair correlations in $A=100$ nuclei involved in neutrinoless double $\beta$ decay}

\author{J.~S.~Thomas}
\affiliation{School of Physics and Astronomy, University of Manchester M13 9PL, United Kingdom}
\author{S.~J.~Freeman}
\affiliation{School of Physics and Astronomy, University of Manchester M13 9PL, United Kingdom}
\author{C.~M.~Deibel}
\altaffiliation[Present address: ]{Department of Physics and Astronomy, Louisiana State University, Baton Rouge, Louisiana 70803, USA}
\affiliation{Physics Division, Argonne National Laboratory, Argonne, Illinois 60439, USA}
\affiliation{Joint Institute for Nuclear Astrophysics, Michigan State University, East Lansing, Michigan 48824, USA}
\author{T.~Faestermann}
\affiliation{Physik Department E12, Technische Universit\"{a}t M\"{u}nchen, D-85748 Garching, Germany}
\affiliation{Maier-Leibnitz-Laboratorium der M\"{u}nchner Universit\"{a}ten (MLL), D-85748 Garching, Germany}
\author{R.~Hertenberger}
\affiliation{Maier-Leibnitz-Laboratorium der M\"{u}nchner Universit\"{a}ten (MLL), D-85748 Garching, Germany}
\affiliation{Fakult\"{a}t f\"{u}r Physik, Ludwig-Maximilians-Universit\"{a}t M\"{u}nchen, D-85748 Garching, Germany}
\author{B.~P.~Kay}
\affiliation{Department of Physics, University of York, Heslington, York YO10 5DD, United Kingdom}
\author{S.~A.~McAllister}
\affiliation{School of Physics and Astronomy, University of Manchester M13 9PL, United Kingdom}
\author{A.~J.~Mitchell}
\affiliation{School of Physics and Astronomy, University of Manchester M13 9PL, United Kingdom}
\author{J.~P.~Schiffer}
\affiliation{Physics Division, Argonne National Laboratory, Argonne, Illinois 60439, USA}
\author{D.~K.~Sharp}
\affiliation{School of Physics and Astronomy, University of Manchester M13 9PL, United Kingdom}
\author{H.-F.~Wirth}
\affiliation{Maier-Leibnitz-Laboratorium der M\"{u}nchner Universit\"{a}ten (MLL), D-85748 Garching, Germany}
\affiliation{Fakult\"{a}t f\"{u}r Physik, Ludwig-Maximilians-Universit\"{a}t M\"{u}nchen, D-85748 Garching, Germany}

\date{\today}

\begin{abstract}
The pairing properties of the neutrinoless double beta decay $(0\nu2\beta)$ candidate $^{100}$Mo have been studied, along with its daughter $^{100}$Ru, to provide input for nuclear matrix element calculations relevant to the decay.  The $(p,t)$ two-neutron transfer reaction was measured on nuclei of $^{102,100}$Ru and $^{100,98}$Mo.  The experiment was designed to have particular sensitivity to $0^{+}$ states up to excitation energies of $\sim 3$~MeV with high energy resolution. Measurements were made at two angles and $L=0$ transitions identified by the ratio of yields between the two angles.  For the reactions leading to and from $^{100}$Ru, greater than $95\%$ of the $L=0$  $(p,t)$ strength was in the ground state, but in $^{100}$Mo about $20\%$ was in excited $0^{+}$ states.  The measured $(p,t)$ data, together with existing $(t,p)$ data, suggest that $^{100}$Mo is a shape-transitional nucleus while $^{100}$Ru is closer to the spherical side of that transition.  Theoretical calculations of the $0\nu2\beta$ nuclear matrix element may be complicated by this difference in shape.
\end{abstract}

\pacs{25.40.Hs, 23.40.Hc, 27.60.+j}

\maketitle

If the process of neutrinoless double $\beta$ decay $(0\nu 2\beta)$ were to be observed, neutrinos would be established as their own antiparticles (Majorana particles) and progress could be made toward determining an absolute scale for the neutrino-mass eigenstates \cite{avignone}.  That neutrinos have mass is established by the observation of neutrino-flavor oscillations \cite{SuperK,K2K,SNO}.  However, such work only establishes differences between the squares of the mass eigenstates.  A determination of the lifetime of the $0\nu 2\beta$ decay process would allow access to the absolute mass scale, provided the mechanism responsible for the decay is driven by light Majorana-neutrino exchange \cite{avignone}.  The rate of the $0\nu 2\beta$ decay is sensitive to nuclear-structure inputs, with the half life given by:
\begin{equation}
\frac{1}{T_{1/2}^{0\nu}} = G_{0\nu}(Q_{\beta \beta},Z) \lvert M_{0\nu} \rvert ^{2} \langle m_{\beta \beta} \rangle^{2}.
\end{equation}
Here, $G_{0\nu}(Q_{\beta \beta},Z)$ is a phase-space factor for the emission of the two electrons in the decay; $\langle m_{\beta \beta} \rangle$ is the effective Majorana mass of the electron neutrino,
\begin{equation}
\langle m_{\beta \beta} \rangle \equiv \lvert \sum_{k} m_{k}U_{ek}^{2} \rvert,
\end{equation}
with $m_{k}$ the neutrino mass eigenvalues, and $U_{ek}$ the ``electron'' row of the neutrino mixing matrix; and $M_{0\nu}$ is the nuclear matrix element describing the decay.

A proper understanding of the nuclear matrix element, $M_{0\nu}$, is necessary if fundamental questions of the properties of neutrinos are to be answered.  This is a difficult problem as there is no experimental probe, apart from the elusive decay itself, that is directly sensitive to the matrix element.  Theoretical input is a necessity.  The matrix element depends on the contributions from a large number of virtual states in the intermediate nucleus in a wide range of spin (up to $\sim 8\hbar$) and excitation energy (up to $\sim 100$~MeV) \cite{avignone,simkovic}.  Specific nuclear structure in this system therefore might not be so important.  However, the wavefunctions of the parent and daughter must be important, and one aspect characterizing them is their pairing properties.  Theoretical approaches to describe the matrix element usually employ the quasiparticle random-phase approximation (QRPA), incorporating both sums of the virtual intermediate states and general pairing properties of the nuclei involved \cite{avignone,simkovic}.  The introduction of like-particle pairing is accomplished through the use of BCS pair correlations \cite{avignone}.  Analysis of QRPA methods shows the importance of $J^{\pi}=0^{+}$ pairs in the decay with $J^{\pi}\ne 0^{+}$ contributions that are small or have phases that result in cancellations \cite{simkovic}.

In this Brief Report we discuss results relevant to the $0\nu2\beta$ decay candidate $^{100}$Mo.  Molybdenum--100 as a decay candidate has advantages for observing the $0\nu2\beta$ decay that experimental groups are exploiting; the high $Z$ and large $Q$-value ($Q_{\beta\beta}=3.034$~MeV) provide a large phase-space factor $(G_{0\nu}\sim Q_{\beta\beta}^{5})$, enhancing the $0\nu2\beta$ decay rate.  The large energy sum of the electrons, $E_{1}+E_{2}=Q_{\beta\beta}$, places signals above most backgrounds \cite{moon,nemo}.  Following previous work on similar decay candidates in the $A=130$ \cite{bloxham} and $A=76$ \cite{freeman} regions, we here examine the pairing properties of $^{100}$Mo and its daughter nucleus $^{100}$Ru through the use of the $(p,t)$ neutron pair transfer reaction, with higher resolution and across a wider region in excitation when compared to previous studies \cite{taketani,taketani2,sharma}.  The objective was to identify $0^{+}$ final states, and to accurately measure their populating cross sections with high energy resolution.  Any significant \textit{differences} between the reactions on $^{100}$Mo and $^{100}$Ru would indicate different pairing properties of the nuclei connected through the $0\nu 2\beta$ decay matrix element, which must be accounted for in theoretical calculations.  The $(p,t)$ reaction was also measured on a target of $^{102}$Ru as the ground state of $^{100}$Ru is populated, and a target of $^{98}$Mo serving as a consistency check.
 
The $(p,t)$ reaction was measured on four isotopically enriched targets of $^{100}$Mo $(97.39\%)$, $^{98}$Mo $(97.18\%)$, $^{100}$Ru $(96.95\%)$, and $^{102}$Ru $(99.38\%)$.  The proton beam was delivered by the MP tandem accelerator of the Maier-Leibnitz-Laboratorium (MLL) of the Ludwig-Maximilians-Universit\"{a}t and Technische Universit\"{a}t M\"{u}nchen at an energy of 24~MeV.  The typical beam current on target was $\sim 450$~nA, and was recorded by a Faraday cup.  The tritons were momentum analyzed using a Q3D magnetic spectrograph.  Separate elastic scattering measurements, at a laboratory angle of $\theta_{\rm lab}=25^{\circ}$, were performed on each target with a 12-MeV $^{3}$He beam to determine the product of target thickness and the solid angle subtended by the spectrograph aperture.  Such a measurement is within the energy regime of Rutherford elastic scattering, and is necessary to convert triton yields from the $(p,t)$ reaction to absolute cross sections.

Charged particles were detected at the focal plane of the spectrometer by a multiwire gas proportional counter backed by a scintillator, providing measurements of focal-plane position, energy loss, and residual energy.  Particle identification was accomplished with the combination of the magnetic-field settings of the spectrograph---the tritons and charged particles from competing reactions have sufficiently different rigidities---and the focal-plane energy signals.  The focal-plane position was determined from the readout of 255 cathode pads on the gas proportional counter.  Each pad has an individual pre-amplifier and shaper, and adjacent pads have a pitch of $3.5$~mm.  A requirement of 3 to 7 adjacent cathode pads with signals above threshold must be met for an event to be registered.  The digitized signals on the active pads were fitted with a Gaussian line shape providing the position measurement to better than $0.1$~mm \cite{wirth}.  

\begin{figure}[ht]
\includegraphics[scale=0.43,clip]{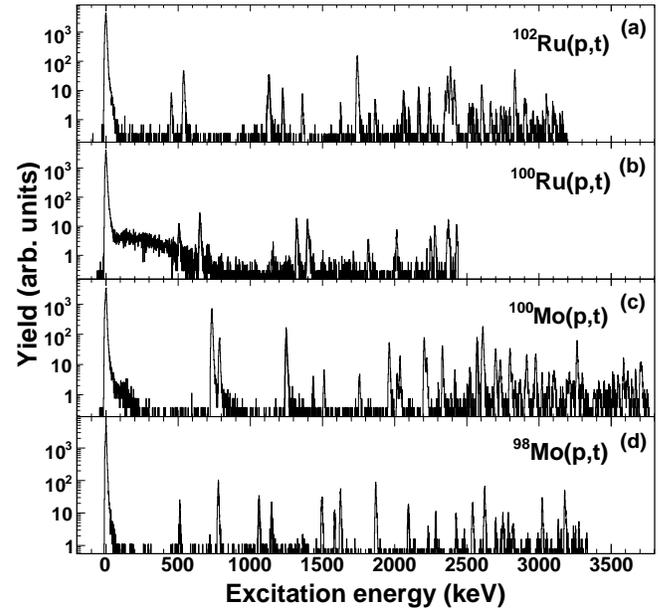}
\caption{\label{yields}The $\theta_{\rm lab}=6^{\circ}$ excitation energy spectra from all spectrograph field settings, combined by normalizing to the relative integrated beams and target thicknesses for, respectively, (a) $^{102}$Ru$(p,t)$, (b) $^{100}$Ru$(p,t)$, (c) $^{100}$Mo$(p,t)$, and (d) $^{98}$Mo$(p,t)$. }
\end{figure}

Triton yields were measured at two spectrograph angle settings ($\theta_{\rm lab}=6^\circ$ and $\theta_{\rm lab}=15^\circ$).  For each target and angle, at least three magnetic-field settings were needed to cover excitation energies up to $E_{x}\sim 3$~MeV.  The focal-plane positions were calibrated to triton momenta with a quadratic polynomial and the excitation energies of known states were reconstructed.  An excitation energy resolution of $\Delta E_{x}\approx 7$~keV was observed.  Care was taken to ensure that suitable overlaps in the corresponding excitation energies between field settings existed.  The resulting excitation energy spectra at $\theta_{\rm lab}=6^{\circ}$ are shown in Fig.\ \ref{yields}.  The triton yields are normalized across the different experimental settings by the corresponding integrated beam currents and target thicknesses.  The background just above the ground states in the spectra for the $^{100}$Ru$(p,t)^{98}$Ru and $^{100}$Mo$(p,t)^{98}$Mo reactions was not identified, but did not hamper the extraction of yields in this excitation energy region.

For even-even nuclei, only the transfer of a pair of nucleons with relative angular momentum $L=0$ is possible to reach $0^{+}$ final states.  Such a transfer is characterized by a forward-peaked angular distribution, at $\theta_{\rm cm}=0^{\circ}$, with all other $L$ transfers peaking at larger angles.  In the BCS model of pairing, nearly all of the pair-transfer strength should be evident in the transition between ground states.  For this reason, and to optimize $L=0$ detection, our measurements were taken as far forward in angle as allowable by focal-plane rate considerations.

\begin{figure}[ht]
\includegraphics[clip,scale=0.49]{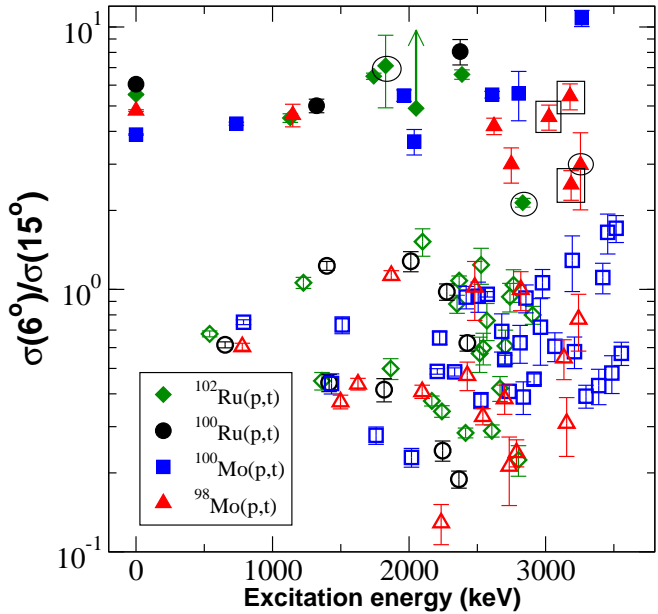}
\caption{\label{ratio}(Color online) The cross section ratios of the states populated in the $(p,t)$ reaction as a function of excitation energy.  The states with ratio larger than 2 (filled symbols) are assigned $J^{\pi}=0^{+}$ in this work. Unfilled symbols are $L > 0$.  Previously unassigned $J^{\pi}$ states are circled, and those assigned a different spin (that may perhaps indicate that the state populated is not the same as in the compilation \cite{nds96,*nds98,*nds100}) are surrounded by a square.}
\end{figure}

The ratio of the yields at $\theta_{\rm lab}=6^{\circ}$ to $\theta_{\rm lab}=15^{\circ}$ is sufficient to distinguish pair transfers of neutrons with relative angular momentum of $L=0$ from those with higher $L$.  Figure \ref{ratio} is a scatter plot of the ratio of measured differential cross sections as a function of excitation energy.  There is a clustering of states involving $L=0$ transfer with $\sigma(6^{\circ})/\sigma(15^{\circ}) > 2$.  In the measured ranges of excitation energy and with this criterion alone, we are able to identify previously assigned $J^{\pi}=0^{+}$ states; confirm six tentatively assigned $0^{+}$ states; make three new $0^{+}$ assignments to previously unassigned states; and alter the assignments of three states that are in conflict with the present results \cite{nds96,*nds98,*nds100}.  Table \ref{summary} summarizes the $J^{\pi}=0^{+}$ states measured in this work, along with the previous information \cite{nds96,*nds98,*nds100}. 

\begin{table*}[ht]
\caption{\label{summary}The states assigned in the present work as $J^{\pi}=0^{+}$.  The previous excitation energies and $J^{\pi}$ assignments are taken from the Nuclear Data Sheets \cite{nds96,*nds98,*nds100}.  There is the possibility where assignments differ from previous work that the state populated here may not be the same as in the compilation.  The measured differential cross sections at $\theta_{\rm lab}=6^{\circ}$ are listed with statistical uncertainties only.  Systematic uncertainties in the absolute cross sections are estimated at $\sim 5\%$.  The relative strength in the last column is computed from the $\theta_{\rm lab}=6^{\circ}$ differential cross section, adjusted for $Q$-value dependence by DWBA calculations, and normalized to the $^{102}$Ru$(p,t)^{100}$Ru, DWBA-adjusted ground state cross section.}
\begin{ruledtabular}
\begin{tabular}{cccccc}
Reaction & $E_{x}$ (keV) current & $E_{x}$ (keV) previous & $J^{\pi}$ previous & $\sigma (6^{\circ})$ (mb/sr) & relative strength \\
\colrule 
$^{102}$Ru$(p,t)^{100}$Ru & 0            & 0        & $0^{+}$ & 4.50(1)                 & $\equiv 100$ \\
                          & $1130.6(3)$  & 1130.317 & $0^{+}$ & $2.72(6)\times 10^{-2}$ & $0.602$ \\
								  & $1742.0(2)$  & 1741.013 & $0^{+}$ & $1.29(2)\times 10^{-1}$ & $2.97$ \\
								  & $1828(2)$    & 1828     & ---     & $1.2(2)\times 10^{-3}$  & $0.028$ \\
								  & $2049(4)$    & 2051.66\footnote{Observed at $\theta_{\rm lab}=6^{\circ}$ but not at $\theta_{\rm lab}=15^{\circ}$. The ratio plotted in Fig.\ \ref{ratio} is therefore a lower limit.}
								             & $0^{+}$ & $4.7(13)\times 10^{-4}$ & $0.011$ \\
								  & $2388.3(3)$  & 2387.38  & $0^{+}$ & $5.0(1)\times 10^{-2}$  & $1.2$ \\
								  & $2833.1(3)$  & 2832.8   & ---     & $3.5(1)\times 10^{-2}$  & $0.93$ \\
$^{100}$Ru$(p,t)^{98}$Ru  & 0            & 0		    & $0^{+}$ & 4.15(1)                 & $85.1$ \\
                          & $1322.1(6)$  & 1322.14  & $0^{+}$ & $1.53(5)\times 10^{-2}$ & $0.353$\\
                          & $2373.9(8)$  & 2374.5   & $(0^{+}$ to $4^{+})$ & $1.30(6)\times 10^{-2}$ & $0.359$\\
$^{100}$Mo$(p,t)^{98}$Mo  & 0            & 0		    & $0^{+}$ & 3.44(1)                 & $79.4$ \\
                          & $734.6(9)$   & 734.75   & $0^{+}$ & $6.43(5)\times 10^{-1}$ & $13.5$\\
                          & $1962.3(3)$  & 1963.08  & $0^{+}$ & $4.57(9)\times 10^{-2}$ & $0.883$\\
                          & $2034.7(5)$  & 2037.53  & $(0^{+},1^{+},2^{+})$ & $1.37(5)\times 10^{-2}$ & $0.264$\\
                          & $2611.3(2)$  & 2608.4   & $0^{+}$ & $1.65(3)\times 10^{-1}$ & $3.17$\\
                          & $2799.6(5)$  & 2803     & $(0^{+})$ & $2.5(1)\times 10^{-2}$ & $0.49$\\
                          & $3264.9(5)$  & 3265     & $(0^{+})$ & $4.8(1)\times 10^{-2}$ & $0.94$\\
$^{98}$Mo$(p,t)^{96}$Mo   & 0            & 0		    & $0^{+}$ & 4.17(1)                  & $78.7$ \\
                          & $1148.0(8)$  & 1148.13  & $0^{+}$ & $1.27(6)\times 10^{-2}$  & $0.228$\\
                          & $2624.2(6)$  & 2622.51  & $(0^{+})$ & $4.8(2)\times 10^{-2}$ & $0.91$\\
                          & $2751(1)$    & 2748.65\footnote{The Nuclear Data Sheets list $0^{+}$ states at two energies, $2742$~keV and $2748.65(7)$~keV \cite{nds96,*nds98,*nds100}. Reference \cite{mo98}, from which the adopted $2742$~keV arises, lists an energy of $2.75$~MeV.}  
								             & $(0^{+})$ & $8.6(8)\times 10^{-3}$ & $0.16$\\
                          & $3023.9(8)$  & 3024.58  & $2^{+}$ & $2.2(1)\times 10^{-2}$   & $0.43$\\
                          & $3178.9(8)$  & 3178.69  & $3^{-}$ & $2.3(1)\times 10^{-2}$   & $0.47$\\
                          & $3185(1)$    & 3186.81  & $4^{+}$ & $1.1(1)\times 10^{-2}$   & $0.22$\\
                          & $3255(2)$    & 3255.63  & ---     & $1.9(4)\times 10^{-3}$   & $0.039$\\
\end{tabular}
\end{ruledtabular}
\end{table*}

The largest $\theta_{\rm lab}=6^{\circ}$ yields are seen for populating ground states for all targets. In the $(p,t)$ reactions on $^{102}$Ru, $^{100}$Ru, and $^{98}$Mo, the next largest yields to $0^{+}$ states are no more than $3\%$ of the respective ground-state yields.  The $(p,t)$ yield from the $^{100}$Mo target is more fragmented, with yields at $6^{\circ}$ of $19\%$ ($E_{x}=735$~keV) and $5\%$ ($E_{x}=2608$~keV) of the ground-state transition.

There is a significant $Q$-value dependence on the cross sections for $(p,t)$ reactions.  To account for this effect in strength comparisons, the measured cross sections were divided by DWBA calculations at the corresponding excitation energy.  The DWBA calculations were performed in a simple, simultaneous-transfer model of the reaction with the code \textsc{ptolemy} \cite{ptolemy}. Global optical-model parameters were used for both the protons \cite{unc} and tritons \cite{pang}.  The $L=0$ neutron pair is bound to the proton, or target-like core, with an energy equal to the respective two-neutron separation energy.  For the purposes of accounting for the $Q$-value dependence, the configuration of the neutron pair is chosen such that the ``bound-state'' form factor has an appropriate number of nodes consistent with pair removal from the $sdg$ shell---only the binding energy changes between form factors for different $Q$-values, not the pairing configuration.

The last column of Table \ref{summary} shows the relative strength of each transition compared to the ground-state transition with the $^{102}$Ru target, after accounting for the $Q$-value dependence with the DWBA calculations.  The three targets of $^{102}$Ru, $^{100}$Ru, and $^{98}$Mo are consistent with $\gtrsim 95\%$ of the $(p,t)$ strength contained in the transfer between ground states.  In contrast, the ground-state transfer is only $\approx 80\%$ of the $(p,t)$ strength to the observed $0^{+}$ states with the $^{100}$Mo target.   

The $A\sim 100$ region near Zr is well-known for a dramatic shape change at $N=60$ \cite{pc}.  However, in the molybdenum isotopes the change is more gradual, as evidenced by changes in nuclear charge radii \cite{charlwood}.  The transition near $^{100}$Mo is characterized by shape-coexistence behavior, with particular consequences for the population patterns observed in $(p,t)$ and $(t,p)$ reactions.

The $(t,p)$ strength for $L=0$ transitions from $^{96}$Mo to states in $^{98}$Mo is concentrated in the ground state, with only $1.5\%$ and $10\%$ fragments of this strength found in the states at $E_{x}=735$~keV and $E_{x}=2608$~keV, respectively \cite{flynn,flynn2}.  In $(t,p)$ reactions to $^{100}$Mo and $^{102}$Mo, it has been observed that an excited $0^{+}$ state near $700$~keV carries $\approx 20\%-30\%$ of the strength of the ground-state transitions \cite{flynn,rahman}.  Larger fragmentation of the $(p,t)$ strength is observed in the current high-resolution measurement with the $^{100}$Mo target, in common with previous studies \cite{taketani, sharma}.  The population patterns of $(p,t)$ and $(t,p)$ reactions are consistent with the transitional nature of the region, and can be understood as the reactions are strong between states of similar deformation.  Both $^{96}$Mo and $^{98}$Mo ground states are approximately spherical in this picture, but the first-excited $0^{+}$ state of $^{98}$Mo could be deformed.  If the $^{100}$Mo ground-state wavefunction contains amplitudes for both spherical and deformed states, a strong overlap is possible with the $0^{+}$ excited state in a $(p,t)$ reaction.  Likewise, the $A > 100$ molybdenum nuclei are gradually more deformed, splitting the $(t,p)$ strength to more states than just the ground state.  A similar situation exists in the Sm isotopes with the $(p,t)$ reaction on the transitional nucleus $^{150}$Sm leading to an excited $0^{+}$ state with $58\%$ the strength of the ground state transition, while the corresponding $^{148}$Sm$(t,p)^{150}$Sm reaction does not populate that state \cite{debenham}.  

The ruthenium isotopes also undergo a gradual shape change near $N=60$.  Measurements of the $^{102,104}$Ru$(t,p)$ reactions were previously performed to investigate the onset of deformation in these nuclei \cite{casten}.  The transitions to the $0^{+}$ states near $990$~keV in both $^{104}$Ru and $^{106}$Ru carry $\approx 20\%$ of strength of the ground-state transitions \cite{casten}.  The $(p,t)$ reactions on targets of $^{100,102,104}$Ru were previously measured with limited energy resolution ($\Delta E_{x}\approx 100$~keV) and excitation energy range (up to $E_{x}=1.5$~MeV) \cite{taketani2}.  An excited $0^{+}$ state was observed at $944$~keV in $^{102}$Ru with $\approx 8\%$ of the ground state strength---the largest such fragment across the measured isotopes \cite{taketani2}.  The present high-resolution measurements on $^{102,100}$Ru targets showed no fragmentation of the two-neutron removal strength to excited $0^{+}$ states larger than $\approx 3\%$ of the ground state transition, up to an excitation energy of $E_{x}\sim 2.5$~MeV.  Summarizing these studies, an excited $0^{+}$ state that is likely associated with the onset of deformation attracts an increasing proportion of the pair-addition strength in reactions on the heavier stable targets.  But for $^{100}$Ru, although several excited $0^{+}$ states are identified, none carry significant strength compared to the ground state. 

The transitional nature of the region around $^{100}$Mo is likely to complicate calculations of double beta decay.  It is known that \textit{differences} in deformation between the parent and daughter nuclei in $0\nu 2\beta$ decay do have a large effect on the calculated QRPA nuclear matrix elements \cite{simkovic_def,fang}.  However, ground-state shapes in the transitional region are also likely to be ill defined with large zero-point fluctuations leading to additional complications.

Returning to the issue of pairing properties, it is noted that the cross sections for population of the ground states in $^{96}$Mo and $^{98,100}$Ru have the same magnitude within $10\%$.  Indeed, the sum of cross sections to the ground and $E_{x}=735$~keV states in $^{98}$Mo is very similar to these ground-state transitions.  This would suggest that, apart from effects of the onset of deformation, the pairing properties of $^{100}$Mo and $^{100}$Ru are broadly similar.

Some data exist on the proton-pair transfer reaction $(^{3}$He$,n)$ on stable $A\sim 100$ targets \cite{fielding}.  Within the experimental sensitivity, no excited $0^{+}$ states were observed in reactions on $^{100}$Mo and $^{100}$Ru, and ground-state reaction cross sections display the same similarity.

In summary, the transitional nature of the region of nuclei around $^{100}$Mo is well known and influences the results of pair transfer studies.  The differences in the extent of deformation between the double beta decay candidate $^{100}$Mo and its daughter $^{100}$Ru, and mixing between different shapes in each nucleus is likely to complicate calculations of matrix elements.  Beyond the effects of the shape change at $N=60$, proton and neutron pairing properties appear to be uncomplicated with no evidence for effects such as pairing vibrations associated with gaps in the underlying single-particle levels.

A summary of all cross section data is available online in the Experimental Unevaluated Nuclear Data List (XUNDL) database \cite{xundl}.
    
\begin{acknowledgments}
The authors would like to acknowledge the operating staff of the Munich tandem.  This work was supported by the UK Science and Technology Facilities Council; the US Department of Energy, Office of Nuclear Physics, under Contract No.\ DE-AC02-06CH11357 and NSF Grant No.\ PHY-08022648; and the DFG Cluster of Excellence ``Origin and Structure of the Universe.'' 
\end{acknowledgments}

\providecommand{\noopsort}[1]{}\providecommand{\singleletter}[1]{#1}%

\end{document}